# HIDDEN SUPERCONDUCTIVITY IN SOLID SOLUTIONS $Sr_{1-x}La_xPbO_3$ DETECTED BY TUNNELING

*Vadim A. Drozd[1*], Toshikazu Ekino[2†], Alexander M. Gabovich[3‡], Marek Pękała[4§], Raquel A. Ribeiro[2**]*

[1] Department of Chemistry, Kiev State University, 64 Volodymyrs'ka st, 01033, Kiev, Ukraine
[2] Hiroshima University, Faculty of Integrated Arts and Sciences, 1-7-1 Kagamiyama, Higashi-Hiroshima, 739-8521, Japan
[3] Crystal Physics Department, Institute of Physics, prospekt Nauki 46, 03028, Kiev-28, Ukraine
Department of Chemistry, University of Warsaw, Al. Żwirki i Wigury 101, PL-02-089 Warsaw, Poland
[4] Department of Chemistry, University of Warsaw, Al. Żwirki i Wigury 101, PL-02-089 Warsaw, Poland

**ABSTRACT**

Solid solutions of $Sr_{1-x}La_xPbO_{3-\delta}$ ($x$ = 0, 0.05, 0.10, 0.15 and 0.20) were synthesized with the help of high pressure. Temperature ($T$) dependences of the resistivity ρ and differential thermoelectric power $S$ were measured. All samples demonstrate an increase of ρ for low $T$ and no signs of a bulk superconducting transition. The sign of $S(T)$ is positive for $x$ = 0.05 and 0.10, whereas $S(T)$ changes sign from positive to negative for $x$ = 0.15. Tunnel measurements were carried out using the break-junction technique. The sample $SrPbO_{3-\delta}$ reveals a correlation dielectric gap Σ ≈ 1.5 ~ 2 eV. On the other hand, $Sr_{0.9}La_{0.1}PbO_{3-\delta}$ shows a Josephson tunnel current and a superconducting gap Δ of approximately 10 meV. This is the first such observation for the studied family of oxides. The results may help to resolve the existing controversy over superconductivity in the conglomerates $Cu_{24}Pb_2Sr_2Ag_2O_x$.

[*] E-mail address: vdrozd@univ.kiev.ua
[†] E-mail address: ekino@hiroshima-u.ac.jp; corresponding author
[‡] E-mail address: gabovich@iop.kiev.ua
[§] E-mail address: pekala@chem.uw.edu.pl
[**] E-mail address: ribeiro@hiroshima-u.ac.jp

1. **Introduction**.

The search for high-critical-temperature superconductors has long become a favorite pursuit of solid-state scientists[1] due to both an apparent technological importance of any significant success[2,3] and a strong intellectual challenge[4,5,6,7]. Oxide ceramics are widely recognized as the most promising group for that purpose[8,9], because their representatives usually have large critical temperatures, $T_c$'s, for relatively low and hence changeable current-carrier densities[8,10], let alone the actual champions – high-$T_c$ cuprate superconductors[11,12]. On the other hand, it is well known that, unfortunately, no theoretically justified first-principle criteria of superconductivity have been established, let alone reliable *a priori* calculations of any critical parameters. Moreover, although electron-phonon interaction is unavoidable in any solid-state object, other superconducting mechanisms (e.g., magnetic or electronic ones) are not ruled out as well[5,6,7,13].

Therefore, certain empirical criteria have been elaborated to explain a variety of superconductors and make at least qualitative predictions of their properties[14,15,16,17]. In particular, for such complex substances as oxides simple heuristic considerations based on the available experience led to a noticeable success in synthesizing new classes of superconductors with interesting properties[18,19,20,21,22].

There are, however, a number of cases where superconductivity was claimed to reveal itself at high temperatures, $T$, as diamagnetic or resistive anomalies. Thus, interface superconductivity was observed both magnetically and resistively at about 91 K in the bronze $Na_xWO_3$[23]. Later on diamagnetic and resistive anomalies were disclosed at even higher temperature 150 K in the bulk of $Na_xWO_3$ single crystals[24]. Hints of room temperature superconductivity were found some time ago for $Cu_{24}Pb_2Sr_2Ag_2O_x$ ceramics[25] and the same group insisted on an observation of a similar phenomenon for the oxide $Ag_xPb_6CO_{9+\beta}$ using both resistive and magnetic probes[26]. Manifestations of superconducting-like resistive anomalies at 190 K $\leq T \leq$ 270 K for the former compound $Cu_{24}Pb_2Sr_2Ag_2O_x$ were subsequently confirmed[27]. It is important to stress that a true superconducting composition (or compositions) in the studied complex conglomerates was not unequivocally identified. Instead, percolation effects and electrical conductivity governed by a weak or strong localization, depending on the $Ag_2O$ content, were searched out in subsequent investigations of the related composite $PbO_2 - xAg_2O - 0.75C$ [28].

To clear up the situation we started a series of syntheses and physical studies of various solid solutions based on the semi-metallic (at high $T$) $SrPbO_3$, its bulk granular specimens being on the verge of the metal-insulator transition, since the electrical resistivity ρ of polycrystals becomes slightly decreasing with $T$ at low $T$[29]. We have shown that substitution of potassium for strontium (nominally hole-like doping) makes the samples even less metallic than the starting ceramics[30], whereas the substitution of lanthanum (nominally electron-like doping) leads to the metallic behavior for all attained temperatures[31], in agreement with previous experiments carried out for smaller doping levels[32]. Still, superconductivity above 20 K was not discovered in our measurements, though we can not exclude it to happen at lower $T$. These results mean that true bulk high-$T_c$ superconductivity claimed to exist in Pb-containing oxides[25,26,27], most probably, is not associated with the phases studied in publications[30,31].

Nevertheless, in this paper we used a powerful local-probe break-junction technique to detect minority superconducting domains in $Sr_{1-x}La_xPbO_{3-\delta}$, if any, i.e. suggesting the ceramics samples to constitute percolative structures[33,34,35], the apparent conductivity of which may be either metallic- or semiconductor-like. Our expectations were realized and superconducting zero-voltage Josephson peaks of the dynamical tunnel conductances $dI/dV$ were found for several samples with $x = 0.1$. Here $I$ is the current and $V$ is the voltage. The same current-voltage characteristics demonstrate also a double-gap structure. It is natural to assume that at least one of the gaps is a superconducting one. We have also measured I-V characteristics for a basic oxide $SrPbO_{3-\delta}$. In addition, in order to analyze the nature of the current carriers we investigated $T$-dependences of the resistivity $\rho$ and the thermoelectric power $S$ for various doping levels $x$.

## 2. Synthesis and sample characterization

Polycrystalline samples of $Sr_{1-x}La_xPbO_{3-\delta}$ ($x$ = 0, 0.05, 0.10, 0.15 and 0.20) were prepared at ambient pressure by a solid state reaction of stoichiometric amounts of $La_2O_3$, $Sr(NO_3)_2$ and PbO. Lanthanum oxide was calcined at 900°C before use to remove hydroxide and carbonate impurities. Powders were ground in an agate mortar and annealed at 750°C in air for 48 hours with intermediate grinding. We used high-pressure synthesis conditions to improve the structural quality of samples, i.e. to reduce the role of granularity, although in the case of polycrystalline ceramics some influence of macro-structural effects is unavoidable[33,34,35]. High-pressure (HP) samples were obtained from the ambient-pressure phases in a cubic anvil press under 7.5 GPa at 400 °C with several minutes of exposure. Oxygen content in solid solutions was determined by the iodometric titration.

The X-ray investigations were carried out by the powder method using the diffractometer DRON-3 (Cu $K_\alpha$- radiation). According to the X-ray diffraction analysis, the single-phase region for the high-pressure $Sr_{1-x}La_xPbO_{3-\delta}$ solid solutions is $0 \leq x \leq 0.15$, which is somewhat broader than for ambient pressure samples ($0 \leq x < 0.15$). All XRD patterns of $Sr_{1-x}La_xPbO_{3-\delta}$ solid solutions were indexed assuming orthorhombically distorted perovskite-type structure.

The orthorhombic deformation of the perovskite structure, $D$, which is an indicator of a crystal structure deviation from the ideal cubic perovskite structure, can be calculated by a formula

$$D = \frac{1}{3}\sum_{i=1}^{3}\left|\frac{a_i - \bar{a}}{a}\right| \cdot 100. \qquad (1)$$

Here $a_i$ denotes crystal lattice parameters $a_1$ = a, $a_2$ = b, $a_3$ = c/√2 and $\bar{a} = \sqrt[3]{a_1 a_2 a_3}$. It is worth noting that orthorhombic distortion of the perovskite structure for all high-pressure samples is smaller than that for ambient pressure samples except for $Sr_{0.95}La_{0.05}PbO_{3-\delta}$. The lowest value of $D$ was observed for $La_{0.9}Sr_{0.1}PbO_{3-\delta}$ (see Table 1).

## 3. Experimental.

Electrical resistivity $\rho(T)$ and thermoelectric power $S(T)$ were studied by the same technique as described in refs. 30,31. The latter quantity was measured between 20 and 300 K in a closed cycle refrigerator, whereas the former one was investigated in the interval 4.2 K ≤ T ≤ 300 K. Tunnel measurements were carried out by the four-probe *in situ* break-junction technique, which had been proved successful to reveal both superconducting[36] and correlation[37] energy gaps.

## 4. Results and discussion.

Dependences $\rho(T)$ measured for compacted HP samples of $Sr_{1-x}La_xPbO_{3-\delta}$ with $0 \leq x \leq 0.2$ are shown in Fig. 1. They should be compared with $\rho(T)$ found for the same series of solid solutions produced at the ambient pressure[31]. In both cases the composition with $x$ = 0.05 has the lowest resistivity among all the compositions studied. The absolute values of the resistivity for $x$ = 0.05 at medium and high $T$ is similar for two kinds of preparation. However, the undoped HP sample has larger magnitude of $\rho$. At the same time, in agreement with our expectations, HP samples with $x$ = 0.15 and 0.2 are an order of magnitude less resistive than their counterparts from ref. 31. Contrary to what was observed earlier[31], now $\rho(T)$ for $x$ = 0.2 lies higher than that for $x = 0.15$. Trying to explain those and other distinctions (see below) one should note that samples synthesized in this work differ from solid solutions investigated in ref. 31 in not only the pressure treatment but also in the way of synthesis.

As for the $T$-dependences of $\rho$, they are quite different for two groups of samples discussed. Specifically, HP specimen of the pure plumbate of strontium $SrPbO_{3-\delta}$ no longer possesses a shallow minimum found previously[30,31,32] in the temperature range 100 K $< T <$ 150 K. But the main new feature of the HP samples is a semiconductor-like temperature behavior for *all studied compositions* (as opposed to the metallic dependences demonstrated by less compacted samples[31]). This seems rather surprising in view of the applied treatment designed to improve metallic trends appropriate to the La-doped $SrPbO_{3-\delta}$.

The observed semiconductive behavior of $\rho(T)$ for ceramics $Sr_{1-x}La_xPbO_{3-\delta}$ might have been explained by the granular nature of the studied structures[33,34,35,38]. However, granularity does not appear to be of the first importance since another set of solid solutions[31] of the same nominal compositions shows metallic character. Hence, our $\rho(T)$ may be considered an intrinsic bulk feature. The three-dimensionality of the investigated ceramics also makes unlikely a detectable role of quantum corrections due to weak-localization and Coulomb interaction effects[39] in determining the electrical conductivity. On the other hand, strong localization precursor phenomena on the metallic side in the neighborhood of the metal-insulator transition may be crucial. In this case a classification of solid solutions according to the sign of the $d\rho/dT$ would be misleading[40,41].

Our attempts to fit the curves displayed in Fig. 1 by the Arrhenius law

$$\rho(T) \propto \exp\left(\frac{A}{T}\right), \qquad (2)$$

where $A$ is an energy gap appropriate to a definite activation process, failed. It means that neither a simple one-body electron band theory nor a charge-density-wave (CDW) one describe transport properties of $Sr_{1-x}La_xPbO_{3-\delta}$ at low $T$. Furthermore, all temperature changes of resistivity are gradual, so that there is no phase transition in the strict sense. The only conclusion is that the observed semiconductive dependences $\rho(T)$ are governed by the Coulomb interaction in a disordered state of the nominally metallic doped system[35,42,40].

The behavior of the latter can be understood bearing in mind a peculiar twofold part played by dopants in oxides. It was fully recognized while studying normal and superconducting properties of cuprates[43]. Thus, in our case the very metallic behavior of the doped strontium plumbate is induced by La doping, the same La ions being centers of the Coulomb scattering and, as a consequence, a cause of a possible current carrier localization. The destructive effect of impurities on electrical conductivity might have disappeared for large enough $x$. However, such large doping degrees are unattainable for chemical (structural) reasons[31]. One shouldn't forget the intrinsic oxygen nonstoichiometry[44] as well, which essentially influences transport properties of oxides[8,45].

Coulomb correlations between La donors in the disordered system $Sr_{1-x}La_xPbO_{3-\delta}$, which can not be considered as a conventional metal in the one-body Wilson sense[46], must substantially modify the activation law (2), whatever physical meaning of the constant $A$. Namely, conductivity mechanism becomes a variable-hopping one

$$\rho(T) \propto \exp\left(\frac{T_{vh}}{T}\right)^{1/n}, \qquad (3)$$

where the number $n$ equals to 4 when the electron density of states (DOS) of the doped current carriers is constant (the Mott law[42,47]) and 2 if a so-called Coulomb gap appears (Efros-Shklovskii law[35,42]). In the former case the parameter $T_{vh}$ is inversely related to the DOS, whereas when the DOS at the Fermi level is depleted by the Coulomb interaction the quantity $T_{vh}$ equals to a certain Coulomb energy value. Real transport measurements for doped semiconductors may demonstrate either $n = 4$ or $n = 2$ in different $T$ ranges[35]. In our case we can not determine exactly the exponent $1/n$ from Fig. 1. It is no great surprise because all formulae of the type (3) are approximate and are obtained suggesting the most probable transport processes only. Nevertheless, $n$ is always larger than unity. In granular metals Coulomb interaction also transforms the tunneling inter-grain conductivity, so that the main term is described by the same eq. (3) with $n = 2$[33,35].

We should stress that the apparent localization manifested by eq. (3) does not mean that it concerns *all current carriers*. Some groups of carriers that remain metallic constitute an ordinary Fermi-liquid not revealed by bulk $\rho(T)$ measurements in the specific solid solutions $Sr_{1-x}La_xPbO_{3-\delta}$ being on the verge of the metal-insulator boundary. More subtle techniques are needed to uncover the veil. It is not so in $BaPb_{1-x}Bi_xO_3$[8] and cuprates[48], where both free and localized (gapped) current carriers are relatively easily seen in the normal state, e.g., by electrical and optical means. In the framework of the Bardeen-Cooper-Schrieffer (BCS) paradigm[6] it seems quite reasonable that a more pronounced role of free holes (in the majority of superconducting oxides) or electrons maintains the high $T_c$ itself.

In Fig. 2 the $T$-dependences of $\rho$ for the least resistive samples ($x = 0.15$ and $x = 0.2$) are shown at the usual scale down to the lowest attained temperatures. One sees that there is no trend towards superconductivity and the samples even appear not to show any sign of normal metallic behavior.

To characterize current carriers of the HP phases we have measured dependences on T of the differential thermoelectric power (Seebeck coefficient) $S$. Corresponding data are displayed in Fig. 3. It turns out that values of $S$ are *positive* for all $T$ when $x = 0.05$ and $0.10$. The composition with $x = 0.15$ changes its sign from positive to negative at $T \approx 25$ K and remains negative for all higher $T$. These results are puzzling because $S$ for the samples of the same or close compositions but synthesized in a different way are negative in the whole temperature range[31,32]. At the same time, one should bear in mind that in nominally hole-doped ceramics $Sr_{1-x}K_xPbO_{3-\delta}$[30] the sign of the curves $S(T)$ is also unexpected, although *negative*.

The observed reduction of the Seebeck coefficient magnitudes (see Fig. 3) as doping increases correlates well with the metallic nature of the current carriers, contrary to the negative sign of $d\rho/dT$ in HP samples of $Sr_{1-x}La_xPbO_{3-\delta}$ (Figs. 1 and 2). Nevertheless, a linear proportionality between $S$ and $T/E_F$, required by the basic theory for conventional metals or semimetals[49], does not hold for the dependences shown in Fig. 3. In addition to the increasing tendency the curves $S(T)$ for $x = 0.05$ and $0.10$ demonstrate also two local maxima or inflection points slightly appearing against this main background. The dependence for $x = 0.15$ has a shallow minimum at about 130 K and a maximum at approximately 260 K. Seebeck coefficient does not become zero for any attainable $T$. Since $S = 0$ in the BCS state[50], the infinite superconducting cluster is absent in any of the investigated samples.

All these features clearly indicate that several groups of current carriers are involved, some of them exhibiting properties pertinent to a degenerate Fermi liquid. It means that localization is not total and one may hope to find superconductivity hidden by the overall semiconductive behavior.

To seek for superconductivity we performed tunnel measurements of HP samples using the break-junction technique[36,37]. High resistivity of the samples has been the main obstacle to these studies. Nevertheless, we were fortunate to obtain tunnel conductances $G(V) \equiv dI/dV$ as functions of $V$ for a pure strontium plumbate and a solid solution with $x = 0.1$.

Conductances for $SrPbO_{3-\delta}$ obtained at $T = 4.2$ K are shown in Figs. 4 and 5. They represent two distinctive groups of results. The first one, depicted in Fig. 4, reveals clear-cut gap-like peculiarities. The patterns are somewhat asymmetrical but the main features are observed for both polarities. Notwithstanding the apparent gap edges, $G(V) \neq 0$ inside the gapped voltage interval for all those junctions. It means that localized and itinerant quasiparticle states coexist.

Let us begin with the curve (a) as a typical specimen of the tunnel data. Gap-like structures at $V \approx \pm 3V$ should be identified with twice the dielectric, most probably, correlation gap $\Sigma$, the whole pattern corresponding to the semiconductor-insulator- semiconductor junction. At the same time, weak shoulders at 1.5 ~ 2 V should be associated with the gap $\Sigma$ itself. The appearance of $\pm\Sigma$ features is probably caused by a partial formation of a semiconductor-insulator-normal metal contact in the break-junction area. Such large activation energy 1.5 ~ 2 eV substantially exceeds any possible transport activation energy $A$ inferred from resistive measurements, if one tries to fit them by the eq. (2) in the limited range of temperatures. This circumstance, most possibly, reflects the bulk many-body Coulomb gap effect[35,40] and, at least partially, the Coulomb blockade effect if tunneling links one or two small conducting grains[35,51]. Curve (b) demonstrates the same-type incomplete correlation gap as its counterpart (a) but gap edges are located at $\pm 1$ V. It is unclear whether this $G(V)$ corresponds to a semiconductor-insulator-normal metal or a semiconductor-insulator- semiconductor junction.

Other junctions manifest quite different tunnel properties described in Fig5. Not only a true dielectric gap but also a finite-width pseudogap is absent. Instead a typical Coulomb V-shaped "gap" with a zero-value electron DOS at the Fermi level shows up in the power-law $G(V)$ dependences[35,42]. Such a behavior was observed, e.g., for strongly correlated oxides $LaNi_xCo_{1-x}O_3$ and $Na_xTa_yW_{1-y}O_3$ on the insulating side of the metal-insulator phase boundary[40].

So far we have considered tunnel data for the undoped ceramics $SrPbO_{3-\delta}$. Break-junction conductivities $G(V)$ obtained at $T= 4.2$ K for the La-doped HP samples $Sr_{0.9}La_{0.1}PbO_{3-\delta}$ are depicted in Fig. 6. All of them give a clear-cut evidence of a *superconducting* Josephson current between electrodes manifested as a zero-bias peak, i.e. Cooper pairing exists in dispersed domains, the volume fraction of the superconducting phase being extremely small, so that any bulk measurements show no signs of superconductivity. We think that suspicious phenomena observed earlier[25,26,27] might be true superconducting ones manifested by a weak superconducting network in the percolating media[41,52,53]. During the last decades a lot of such structures deliberately prepared to study phase transitions between macroscopically insulating and macroscopically superconducting structures were investigated[35]. In particular, one can mention amorphous InO films[54,55], Cd-Sb alloys[56], Al-Ge granular films[57] and non-stoichiometric $Mg_{1-x}B_2$ polycrystalline samples[58]. High-$T_c$ cuprates have been also discussed from this viewpoint[41]. One can also imagine a possibility that in our case the detected superconducting phase is restricted to surface areas of the substance. Hence, the observation of superconducting correlations by bulk transport measurements becomes impossible in principle. Surface superconductivity, in its turn, may be connected to specific chemical composition of the grain and crack boundaries or to modified electron structure and/or electron-phonon interaction at the interface[59].

The emergence of superconductivity in $Sr_{0.9}La_{0.1}PbO_{3-\delta}$, manifested by the coherent supercurrent, is also confirmed by the development of the superconducting energy gap in the electron spectrum (see Fig. 6), reflecting the semiconductor-like aspect of the superconductivity phenomenon[49,60]. To be specific, gap-like features include dips at $\pm 10$ meV and shoulders at $\pm 20$ meV. Those shoulders are similar to common broadened superconducting-gap-edge peculiarities observed in many break-junction tunneling spectra of different superconductors. Thus, we can attribute $\pm 20$ meV to the $\pm 2\Delta$ value for a superconductor-insulator-superconductor break junction, whereas $\pm 10$ meV dips (or shoulders just outside the dip) should be regarded as $\pm\Delta$ edges, appearing in a superconductor-insulator-normal metal junction formed inside the break junction. The reason of the dip feature appearance instead of the expected hump remains unclear. Nevertheless, the superconducting nature of all singularities found in our spectra seems beyond question in view of the Josephson central peak and the scale of the feature positions being substantially reduced relative to the correlation gaps observed for $SrPbO_{3-\delta}$ (see Figs.4 and 5).

Taking $2\Delta = 20$ meV and assuming the BCS ratio[49] $2\Delta/ k_B T_c \approx 3.5$ to be valid, we can estimate the $T_c$ value for $Sr_{0.9}La_{0.1}PbO_{3-\delta}$ as about 65 K. It is remarkable that at this very temperature the magnitude of the thermoelectric power $S(T)$ falls slightly but abruptly (see Fig. 3), which means the reduction of the superconducting phase fraction of the sample. For $x = 0.05$ a cusp is seen in $S(T)$ at about 50 K, whereas for $x = 0.15$ the feature point moves to much larger temperatures. Measurements of temperature dependences of the dip-shoulder structures revealed in our tunnel spectra will be our next step in order to elucidate the nature of the phenomena observed.

### 5. Conclusions.

We have carried out transport and tunnel investigations of ceramics $Sr_{1-x}La_xPbO_{3-\delta}$ synthesized with the help of high pressure. Pure $SrPbO_{3-\delta}$ is a genuine insulator with strong manifestations of current carrier localization at low temperatures. Samples of $Sr_{0.9}La_{0.1}PbO_{3-\delta}$ incorporate small superconducting domains. Superconductivity was found only by the break-junction tunneling used as a local probe. We observed both coherent Josephson current and superconducting gap-like features. We estimate the gap $\Delta$ to be approximately 10 meV, so that according the BCS theory $T_c$ should be about 65 K. At the same time, the superconducting clusters are beyond the percolation threshold and did not reveal themselves in bulk resistive and thermoelectric measurements. The whole body of data demonstrates an interesting interplay of localization phenomena and superconductivity. Since the sample properties are extremely sensible to conditions of synthesis, it seems possible that modifications of the preparation method may overcome the localization trends and lead to even higher-$T_c$ superconductivity with a larger bulk fraction of the superconducting phase.

**Acknowledgements**

V.A.D. and A.M.G. are grateful to the Mianowski Foundation for support of their visits to Warsaw University. A.M.G. thanks the Japan Society for the Promotion of Science for support of his visit to the Hiroshima University (Grant ID No. S-03204). The research has been partly supported by the NATO grant PST.CLG.979446 and the Grant-in-Aid for COE Research (No. 13CE2002) and for Scientific Research (No. 15540346) of the Ministry of Education, Culture, Sports, Science and Technology of Japan.

FIGURE CAPTIONS

Fig. 1. Temperature, $T$, dependences of the resistivity $\rho$ for $Sr_{1-x}La_xPbO_{3-\delta}$ with $x = 0$, 0.05, 0.10, 0.15 and 0.20.

Fig. 2. The same as in Fig. 1 for most conducting samples with the emphasis on low-$T$ behavior.

Fig. 3. The $T$ dependences for the differential thermoelectric power $S$ ($x = 0.05$, 0.10 and 0.15).

Fig. 4. The dependences on voltage $V$ of the differential conductivity $G(V) \equiv dI/dV$ for break junctions made of $SrPbO_{3-\delta}$, where $I$ is the tunnel current. The curves $G(V)$ demonstrate finite-width correlation gaps.

Fig. 5. The same as in Fig. The curves $G(V)$ V-shaped Coulomb gaps.

Fig. 6. The same as in Figs. 4 and 5 but for $Sr_{0.9}La_{0.1}PbO_{3-\delta}$.

**Table 1. Orthorhombic crystal lattice parameters and oxygen content for $Sr_{1-x}La_xPbO_{3-\delta}$ solid solutions**

| x | a, Å | b, Å | c, Å | V, Å$^3$ | D,% | y | Pb AV |
|---|---|---|---|---|---|---|---|
| 0 | 5.855(4) | 5.951(2) | 8.319(5) | 289.9(5) | 0.624 | 2.96 | 3.92 |
| 0-HP | 5.822(8) | 5.916(4) | 8.297(8) | 285.8(9) | 0.546 | 2.94 | 3.88 |
| 0.05 | 5.840(7) | 5.950(4) | 8.333(6) | 289.6(7) | 0.637 | 2.94 | 3.83 |
| 0.05-HP | 5.81(2) | 5.912(8) | 8.33(2) | 286.5(1.8) | 0.678 | 2.93 | 3.81 |
| 0.10 | 5.860(5) | 5.952(2) | 8.327(7) | 290.4(6) | 0.591 | 2.92 | 3.74 |
| 0.10-HP | 5.84(3) | 5.92(1) | 8.31(3) | 287.3(2.8) | 0.416 | 2.88 | 3.66 |
| 0.15 | 5.861(7) | 5.960(4) | 8.355(7) | 291.8(8) | 0.572 | 2.90 | 3.65 |
| 0.15-HP | 5.86(3) | 5.926(9) | 8.32(3) | 288.8(2.6) | 0.420 | 2.90 | 3.65 |
| 0.20 | 5.858(3) | 5.955(2) | 8.337(3) | 290.8(3) | 0.594 | 2.89 | 3.58 |
| **0.20-HP** | 5.83(2) | 5.926(7) | 8.37(2) | 287.2(1.8) | 0.574 | 2.89 | 3.58 |

*Pb AV* denotes lead average valence, HP symbolizes high-pressure samples.

Figure 1

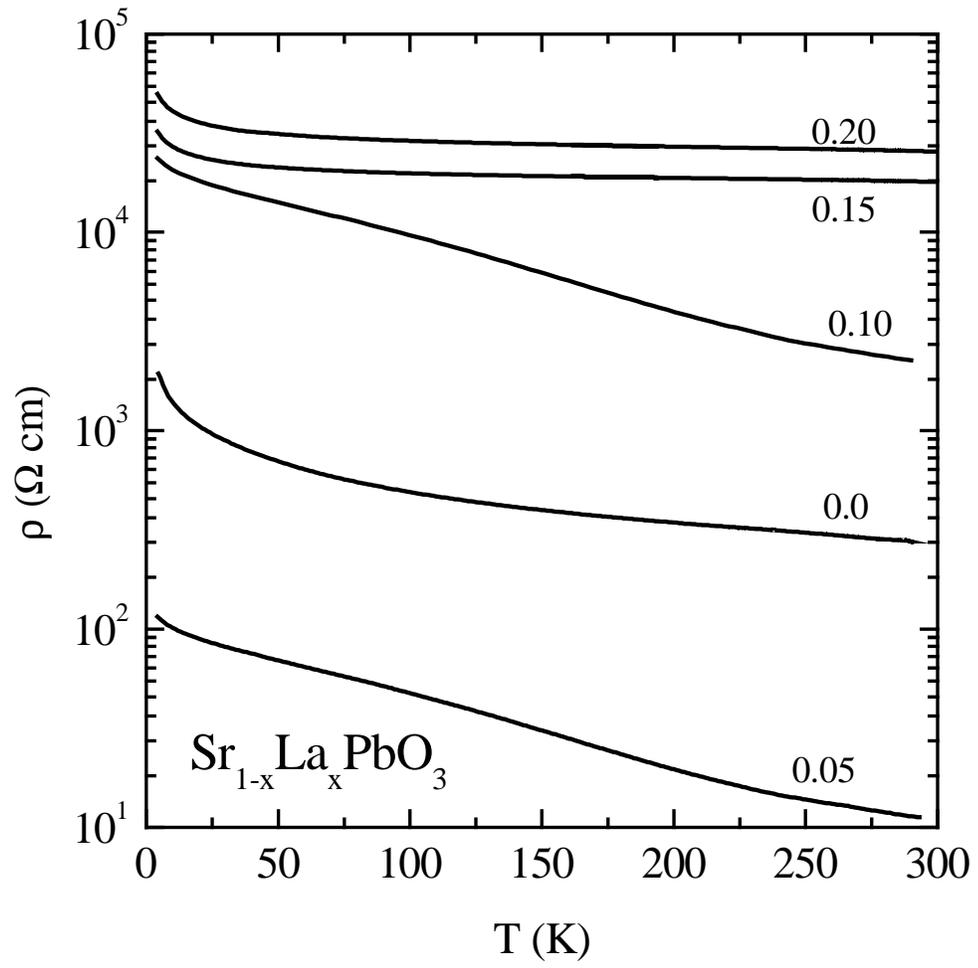

Figure 2

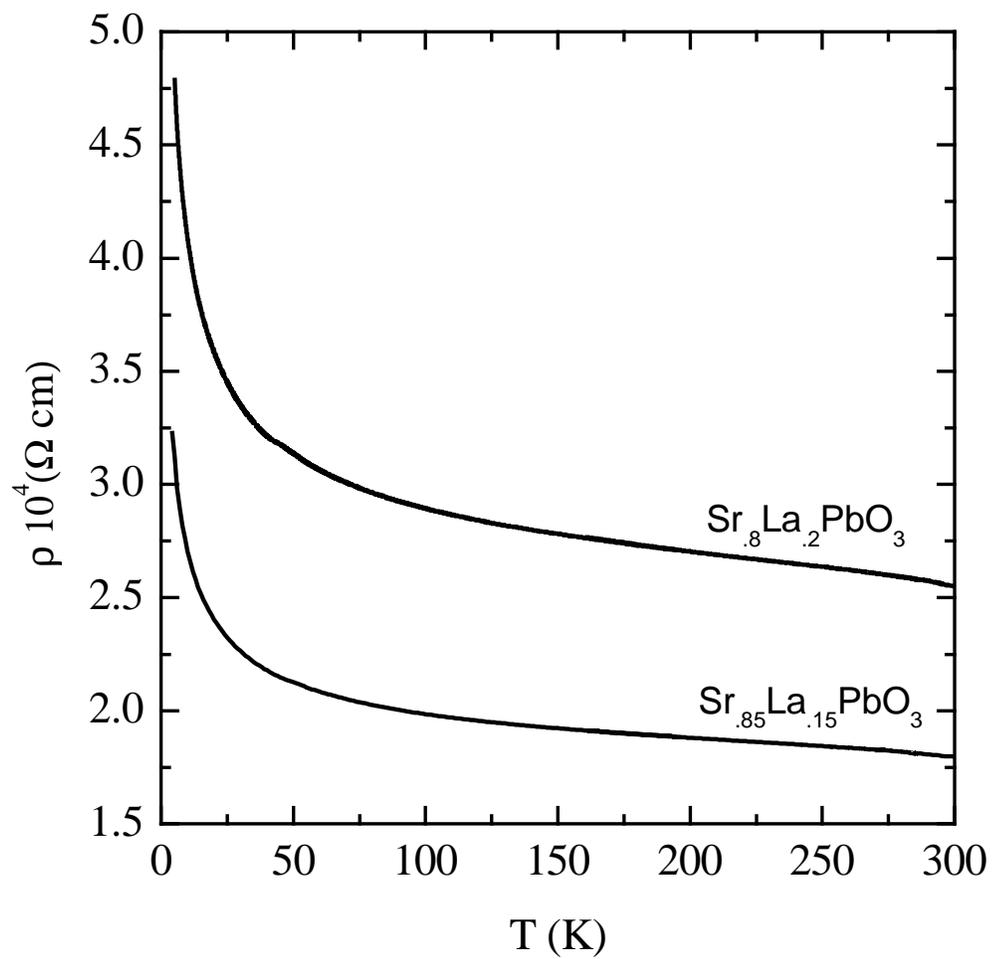

Figure 3

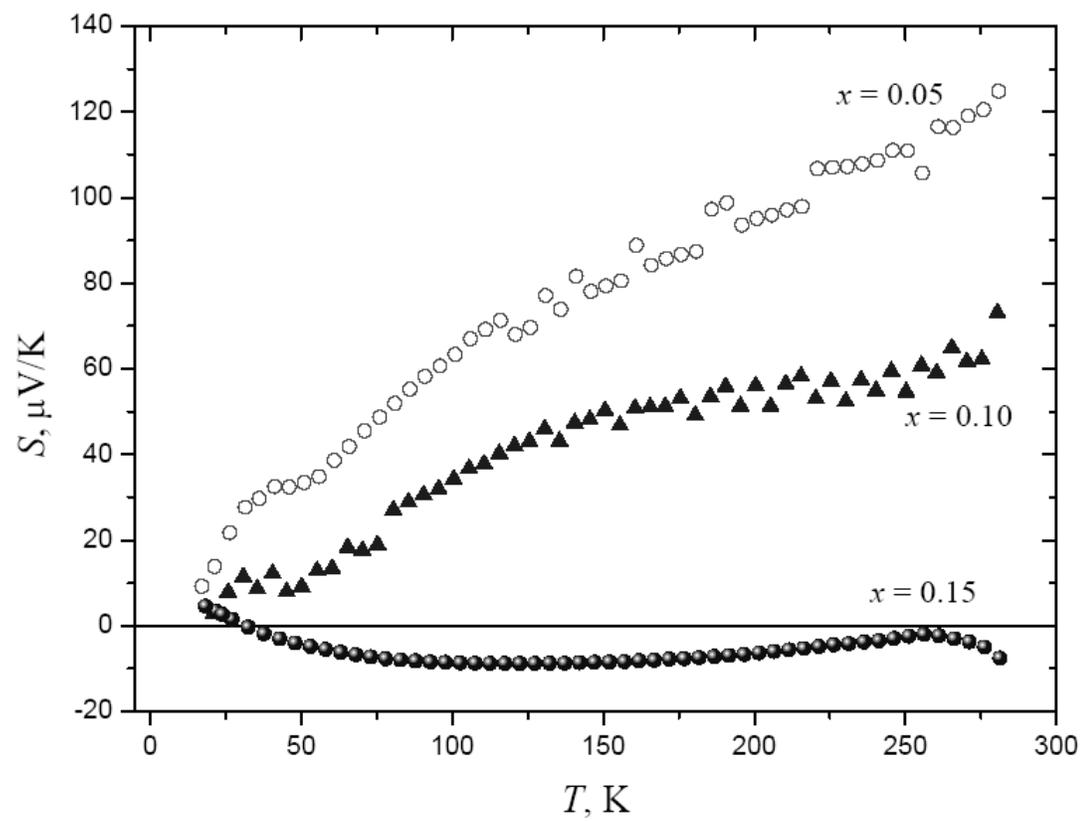

Figure 4

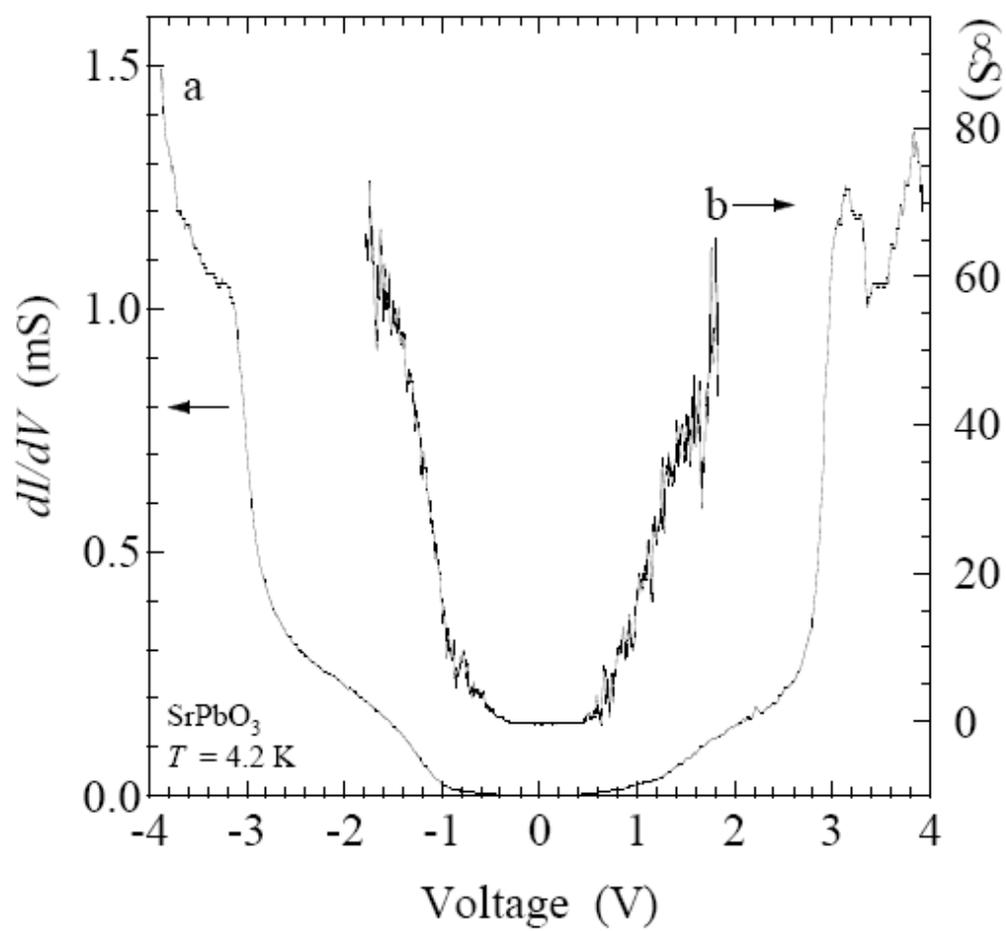

Figure 5

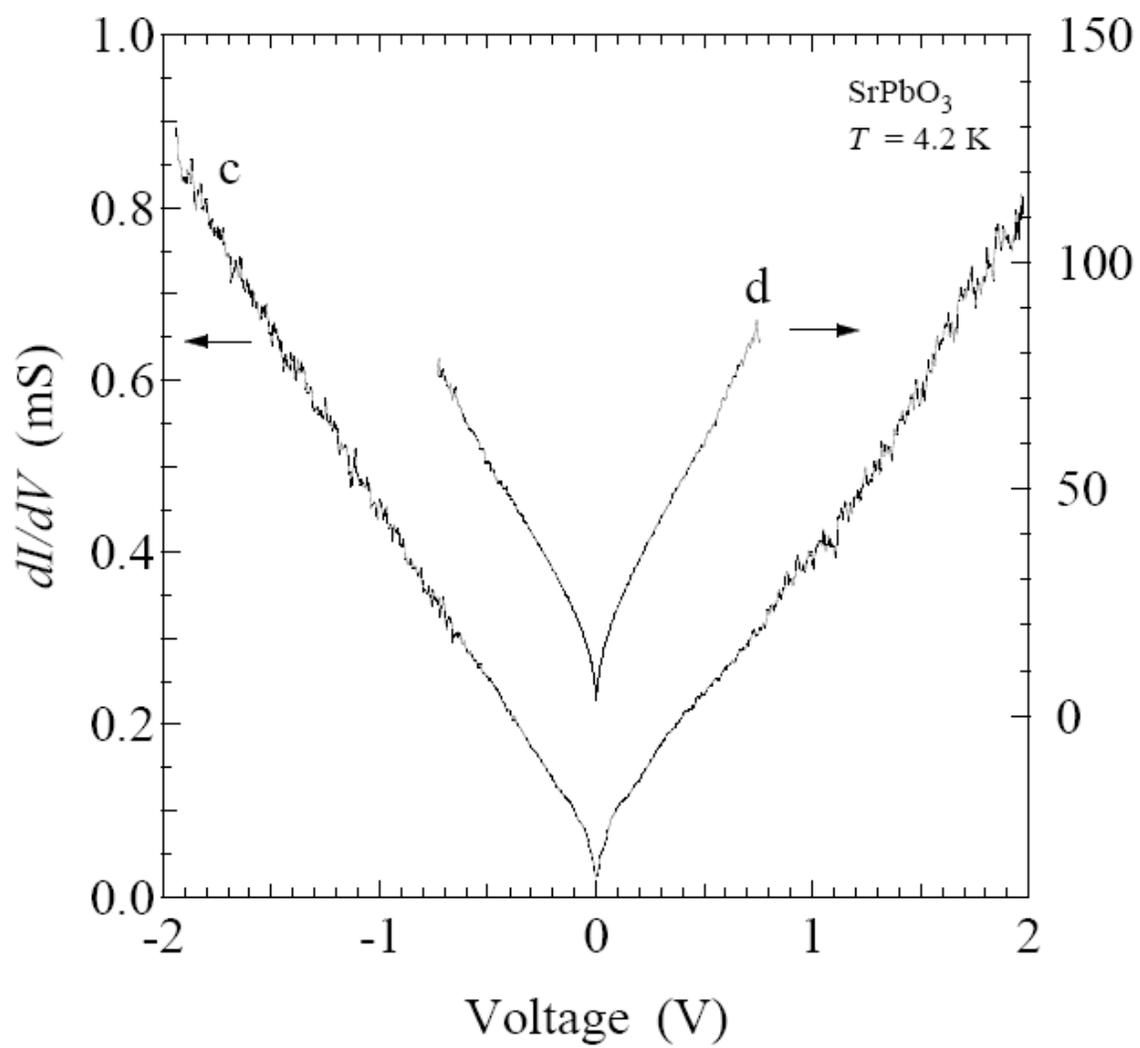

Figure 6

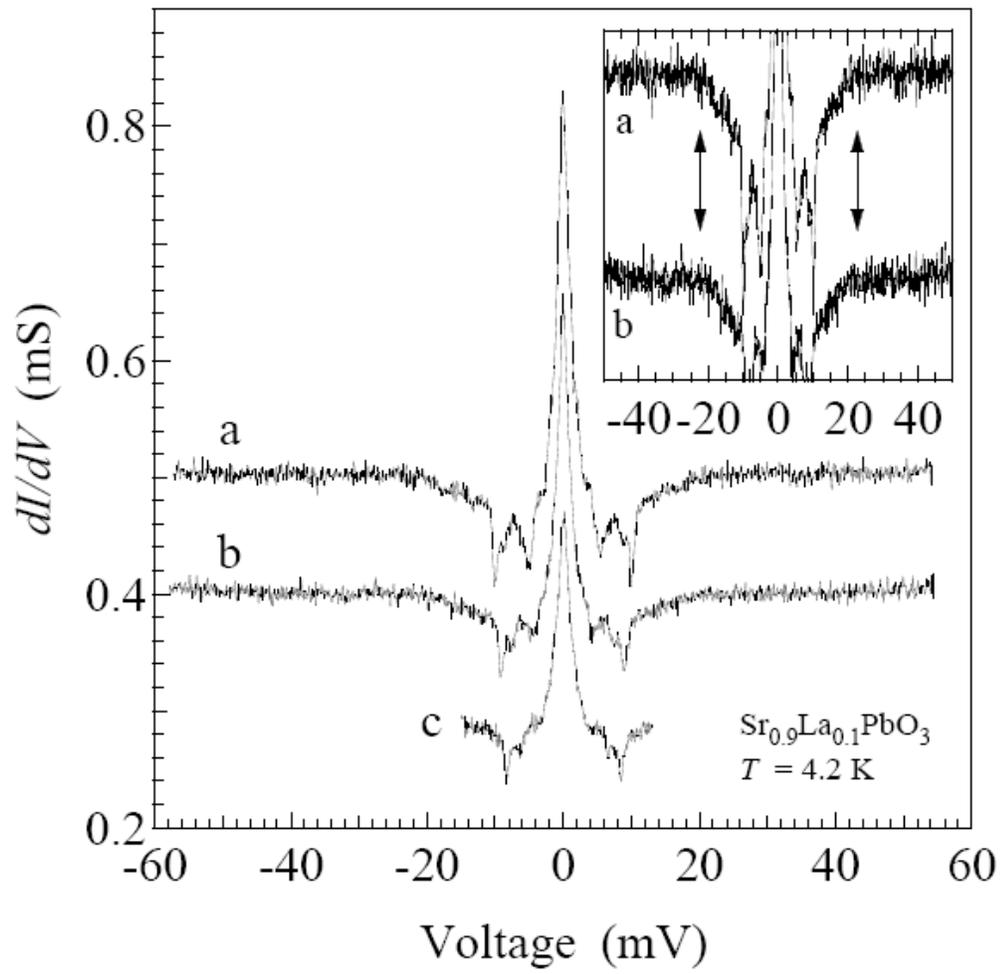